# A Study of the Matter of SPH Application to Saturated Soil Problems

Ha H. Bui, R. Fukagawa
Department of Civil Engineering
Ritsumeikan University
Shiga, Japan
hhbui@fc.ritsumei.ac.jp; fukagawa@se.ritsumei.ac.jp

K. Sako
Global Innovative Research Organization (GIRO)
Ritsumeikan University
Shiga, Japan
kaz-sako@fc.ritsumei.ac.jp

*Abstract*—**We present an application of SPH to saturated soil problems. Herein, the standard SPH formulation was improved to model saturated soil. It is shown that the proposed formulation could yield several advantages such as: it takes into account the pore-water pressure in an accurate manner, it automatically satisfies the dynamics boundary conditions between submerged soil and water, and it reduced the computational cost. Discussions on the use of the standard and the new SPH formulations are also given through some numerical tests. Furthermore, some techniques to obtained correct SPH solution are also proposed and discussed. To the end, this paper suggests that the proposed SPH formulation should be considered as the basic formulation for further developments of SPH for soil-water couple problems.**

## I. INTRODUCTION

The finite element method (FEM) has been employed as the standard numerical approach in computational geomechanics. However, most of problems in geotechnical engineering often involved the large deformation and post-failure problems such as: post-failure process of a sliding slope, debris flow in landslide, seepage failure, post-failure of a slope due to soil liquefactions, etc. In such circumstances, FEM suffers several disadvantages due to mesh tangling even when the updated Lagrangian method is adopted. Re-meshing may help to resolve this problem but the procedure is quite complicated. As an alternative for such computational purposes, it is attractive to develop mesh-free methods. So far, the most popular mesh-free method in geotechnical engineering is the discrete element method (DEM) which tracks motion of a large number of particles, with interparticle contacts modeled by spring and dashpot systems [1]. The main advantages of this approach are that it can handle large deformation and failure problems; and the concept is relatively simple and easy to implement in a computer code. However, DEM suffers from low accuracy since suitable parameters for the contact model are difficult to determine. The discontinuous deformation analysis (DDA) method [2] has also been applied in geotechnical applications, but is mainly used for rock engineering, etc. On the other hand, the method of smoothed particle hydrodynamics (SPH) [3-4] has been recently developed for solving large deformation and post-failure flows of geomaterials [5-13], and represents a powerful way to understand and quantify the failure mechanisms of soil in such challenging problems.

When solving the two-phase water saturated soil, it is the common approach in computational geomechanics to treat the two-phase system as a single phase and the interaction between soil and water was considered via the contribution of the pore-water pressure using the Terzaghi's effective stress concept. Pastor et al. [10] employed this approach in his SPH model to take into account contribution of the pore-water pressure in the landslide simulation. The gradient of pore-water pressure in his model has been approximated using the conventional SPH formulation. However, our current research on the application of SPH to saturated soil revealed that such the approximation of the gradient of the pore-water pressure will lead to numerical instability problem which may failure the SPH computational process for cases when soil is completely submerged into water. Therefore, it is necessary to overcome this limitation in order to generalize the SPH applications to computational geomechanics. In this paper, we will firstly demonstrate the numerical instability problem caused by using the conventional SPH formulation. Then, we will derive a general SPH formulation which can be applied to both dry and saturated soils. Finally, we will show some advantages of the propose SPH formulation. Technique to obtain the initial stress condition of soil in SPH is also proposed by adding a damping force into the motion equation. Several numerical testes are performed to validate the proposed formulation.

## II. SMOOTHED PARTICLE HYDRODYNAMICS

### A. Standard SPH Formulations

In SPH, approximations for quantities of a continuum field such as density, velocity, pressure, etc., are performed using the following interpolation function,

$$A(\mathbf{r}) = \int A(\mathbf{r}')W(|\mathbf{r} - \mathbf{r}'|, h)\, d\mathbf{r}' \qquad (1)$$

where *A* is any variables defined on the spatial coordinate **r**, and *W* is smoothing kernel, which is chosen herein to be the cubic-spline function [14],

$$W(q) = \alpha_d \times \begin{cases} 1 - \frac{3}{2}q^2 + \frac{3}{4}q^3 & 0 \leq q < 1 \\ \frac{1}{4}(2-q)^3 & 1 \leq q < 2 \\ 0 & q \geq 2 \end{cases} \quad (2)$$

where $\alpha_d$ is the normalization factor which is $10/7\pi h^2$ in two-dimensional problems and $q$ is the normalized distance $q=|\mathbf{r}|/h$.

The integral (1) is then discretised onto a finite set of interpolation points (particles) by replacing the integral by a summation and the mass element $\rho V$ with the particle mass $m$,

$$A(\mathbf{r}) \approx \sum_{b=1}^{N} m_b \frac{A_b}{\rho_b} W(|\mathbf{r} - \mathbf{r}_b|, h) \quad (3)$$

where subscript $b$ refers to the quantity evaluated at the position of particle $b$. This "summation approximation" is the basis of all SPH formalisms.

The SPH approximation for the gradient terms may be calculated by taking analytical derivative of equation (3), giving:

$$\nabla A(\mathbf{r}) \approx \sum_{b=1}^{N} m_b \frac{A_b}{\rho_b} \nabla_a W_{ab} \quad (4)$$

where we have assumed that the gradient is evaluated at another particle $a$ ($\mathbf{r} = \mathbf{r}_a$) and the remaining terms are defined,

$$W_{ab} \equiv W(|\mathbf{r}_a - \mathbf{r}_b|, h), \text{ and } \nabla_a W_{ab} \equiv \frac{\mathbf{r}_{ab}}{|\mathbf{r}_{ab}|} \frac{\partial W_{ab}}{\partial \mathbf{r}_a} \quad (5)$$

However, this form of gradient is not guaranteed to vanish when $A(\mathbf{r})$ is constant. To ensure that it does, the gradient can be written as,

$$\nabla A_a \approx \sum_{b=1}^{N} m_b \frac{(A_b - A_a)}{\rho_b} \nabla_a W_{ab} \quad (6)$$

Alternatively, the following forms of the gradient approximations which are the most commonly used to discrete the momentum equation can be written as,

$$\nabla A_a \approx \rho_a \sum_{b=1}^{N} m_b \left( \frac{A_a}{\rho_a^2} + \frac{A_b}{\rho_b^2} \right) \nabla_a W_{ab} \quad (7)$$

$$\nabla A_a \approx \rho_a \sum_{b=1}^{N} m_b \frac{(A_b + A_a)}{\rho_a \rho_b} \nabla_a W_{ab} \quad (8)$$

Further details of SPH literature can be found in [15-16].

### B. Kernel gradient correction

The above SPH gradient approximations may have low accuracy due to the particle deficiency, especially for the region near the boundary surface. In such a case, it is necessary to improve the accuracy of these gradient approximations. Several methods have been proposed to address this issue [17-20]. In this paper, the correction technique [20] will be adopted to improve the accuracy of the gradient approximations. Accordingly, in order to ensure exactly gradient of a linear field, the kernel derivative was normalized in the following manner,

$$\tilde{\nabla} W_{ab} = L(\mathbf{r}) \nabla W_{ab} \quad (9)$$

where $L(\mathbf{r})$ is defined by,

$$L(\mathbf{r}) = \left( \sum_{b=1}^{N} \frac{m_b}{\rho_b} \nabla W_{ab} \otimes (x_a - x_b) \right)^{-1} \quad (10)$$

As a result, all the kernel gradients appeared in equations (6-8) should be replaced by the correction (9).

## III. SPH DISCRETIZATION OF MOTION EQUATION

In this section, discretization of the motion equation of soil using the conventional SPH formulation is presented. Next, a hybrid equation which addresses the numerical instability issue caused by the conventional formulation is derived.

### A. Motion equation

The motion of a continuum can be described through the following equation,

$$\rho \ddot{u}^\alpha = \nabla_\beta \sigma^{\alpha\beta} + \rho g^\alpha \quad (11)$$

where $u$ is the displacement; $\alpha$ and $\beta$ denote Cartesian components $x$, $y$, $z$ with the Einstein convention applied to repeated indices; $\rho$ is the density; $\sigma$ is the total stress tensor, taken negative for compression; and $g$ is the acceleration due to gravity. For a soil, the total stress is normally composed of the effective stress tensor ($\sigma'$) and the pore-water pressure ($p_w$),

$$\sigma^{\alpha\beta} = \sigma'^{\alpha\beta} + p_w \delta^{\alpha\beta} \quad (12)$$

When the pore-water pressure is zero, the displacement $u$ of soil particles relates to the effective stress in the following way,

$$\rho \ddot{u}^\alpha = \nabla_\beta \sigma'^{\alpha\beta} + \rho g^\alpha \quad (13)$$

Using equations (7) and (8), the partial differential form of equation (13) can be approximated in the SPH formulation in the following ways,

$$\ddot{u}_a^\alpha = \sum_{b=1}^{N} m_b \left( \frac{\sigma'^{\alpha\beta}_a}{\rho_a^2} + \frac{\sigma'^{\alpha\beta}_b}{\rho_b^2} + C_{ab}^{\alpha\beta} \right) \tilde{\nabla}_a^\beta W_{ab} + g_a^\alpha \qquad (14)$$

$$\ddot{u}_a^\alpha = \sum_{b=1}^{N} m_b \left( \frac{\sigma'^{\alpha\beta}_a + \sigma'^{\alpha\beta}_b}{\rho_a \rho_b} + C_{ab}^{\alpha\beta} \right) \tilde{\nabla}_a^\beta W_{ab} + g_a^\alpha \qquad (15)$$

where $a$ indicates the particle under consideration; $\rho_a$ and $\rho_b$ are the densities of particle $a$ and $b$ respectively; $N$ is the number of "neighbouring particles", i.e. those in the support domain of particle $a$; $m_b$ is the mass of particle $b$; $C_{ab}^{\alpha\beta}$ is a stabilization term employed to remove the stress fluctuation and tensile instability. The stabilization term consists of two components: artificial viscosity and artificial stress, which could be computed similarly to Bui et al. [7-8] except that the sound speed for the artificial viscosity term is calculated herein by,

$$c_a = \sqrt{G_a / \rho_a} \qquad (16)$$

where $G$ is the shear modulus of soil. Equations (14) and (15) can apply well to dry or single-phase soils and yield no significant difference in computational results for homogeneous ground density. However, when applying to non-homogenous soils and the continuity equation of soil is resolved, equation (14) may encounter difficulties in dealing with density ratio $\rho_1/\rho_2 \leq 0.5$, where $\rho_1$ and $\rho_2$ are the density between two adjacent soil layers. The presence of a sharp density gradient at the interface is the main source of a severe instability problem, and hence alteration form of equation (15) is chosen throughout this paper [21].

### B. An improvement of the motion equation for saturated soil

To account for the pore-water pressure in soil deformation analyses, the effective stress tensor in equation (15) must be replaced by the total stress tensor defined by equation (12). Accordingly, the conventional SPH formulations for saturated soil can be written as follows:

$$\ddot{u}_a^\alpha = \sum_{b=1}^{N} m_b \left( \frac{\sigma'^{\alpha\beta}_a + \sigma'^{\alpha\beta}_b}{\rho_a \rho_b} + C_{ab}^{\alpha\beta} \right) \tilde{\nabla}_a^\beta W_{ab} \qquad (17)$$
$$+ \sum_{b=1}^{N} \frac{m_b}{\rho_a \rho_b} (p_{wb} + p_{wa}) \tilde{\nabla}_a^\alpha W_{ab} + g_a^\alpha$$

Numerical tests, as given later in section V, reveal however that this equation leads to numerical instability, and thence failure of the SPH computational process, for soil particles near an interface between submerged soil and a water reservoir. This numerical instability, in some cases, would result in a problem that particles located near the interface are expelled from the soil structure. We have tried to resolve this problem by imposing water pressure on soil particles on the boundary between submerged soil and water, but no improvement was obtained. Thus, it is necessary to modify this equation for application to saturated soil.

To explain the above numerical instability as well as our modification, let us consider the original SPH discretization of the gradient of the pore-water pressure term that appeared in equation (17),

$$\frac{1}{\rho_a} \int_\Omega (p_{wb} + p_{wa}) \nabla_a W_{ab} dV_b \qquad (18)$$

where $dV_b$ is the volume of particle $b$; $\Omega$ is the support domain which includes all neighbours $b$ of particle $a$. This equation can be further written as,

$$\frac{1}{\rho_a} \int_\Omega (p_{wb} - p_{wa}) \nabla_a W_{ab} dV_b + \frac{2 p_{wa}}{\rho_a} \int_\Omega \nabla_a W_{ab} dV_b \qquad (19)$$

Next, let's consider a special case where soil is submerged into a constant pore-water pressure field as shown in Figure 1. We will employ expression (19) to calculate the gradient of the pore-water pressure of a particle $a$ that is located right on the boundary between submerged soil and water; expression (19) is correct only if this gradient is zero everywhere within the soil domain.

The first term of expression (19) is zero everywhere due to the constant pore-water pressure field assumption. Next, we will prove that the second term is zero everywhere within the submerged soil domain except the area near the ground surface. Using the divergence theorem, the second term in expression (19) can be transformed as follows,

$$\frac{2 p_{wa}}{\rho_a} \int_\Omega \nabla_a W_{ab} dV_b = \frac{2 p_{wa}}{\rho_a} \left( \int_{ACB} W_{ab} \vec{n}_1 ds_b - \int_{AB} W_{ab} \vec{n}_2 ds_b \right) \qquad (20)$$

where $S$ is the surface or edge of $\Omega$; and $\vec{n}$ is the unit vector locally normal to the surface $S$.

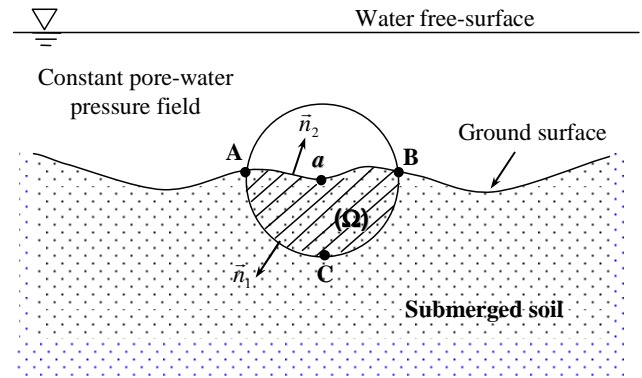

Figure 1.　Soil submerged into a constant pore-wate pressure field.

Since the kernel function $W$ is defined to have compact support, the surface integral on ACB, as shown in Figure 1, is zero. Finally, we end up with,

$$\frac{2p_{wa}}{\rho_a}\int_\Omega \nabla_a W_{ab} dV_b = -\frac{2p_{wa}}{\rho_a}\int_{AB} W_{ab}\vec{n}_2 ds_b \qquad (21)$$

Equation (21) is similar to the pressure integral on the surface AB, which results in a force acting on particle $a$ in the direction of unit vector $n_2$ as the pore-water pressure is negative in the current simulation. Consequently, soil particles that are located near the interface between the submerged soil and water are expelled from the soil if the pore-water pressure on the interface is large enough. In this paper, to avoid this problem, we suggest to remove the second term of expression (19) from the gradient approximation of the pore-water pressure. As a result, the gradient of the pore-water pressure at particle $a$ can be approximated using the following equation,

$$\left(\frac{1}{\rho}\nabla p_w\right)_a = \sum_{b=1}^N \frac{m_b}{\rho_a \rho_b}(p_{wb}-p_{wa})\tilde{\nabla}_a W_{ab} \qquad (22)$$

which ensures that the gradient of a constant pore-water pressure field vanishes. One advantage of using equation (22) is that it automatically satisfies the boundary condition between submerged soil and water, i.e. there is no need to impose water pressure on soil particles at the interface between submerged soil and water. This significantly reduces computational time by avoiding a search for particles on the boundary between water and submerged soil. This automatic achievement of boundary conditions comes from the fact that the pore-water pressure of particle $a$ is subtracted from the pore-water pressure of neighbouring particles in equation (22).

Equation (22) can also be derived using the following transformation,

$$\left(\frac{1}{\rho}\nabla p_w\right)_a = \left(\frac{1}{\rho}[\nabla p_w - p_w \nabla 1]\right)_a$$
$$= \sum_{b=1}^N \frac{m_b}{\rho_a\rho_b}(p_{wb}-p_{wa})\tilde{\nabla}W_{ab} \qquad (23)$$

which is equal to the SPH approximation in equation (22). This suggests that removing the last term in expression (19) does not cause any inconsistency in the gradient approximation of the pore-water pressure. Accordingly, the new SPH equation for the motion of saturated soil is written as follows,

$$\ddot{u}_a^\alpha = \sum_{b=1}^N m_b\left(\frac{\sigma_a'^{\alpha\beta}+\sigma_b'^{\alpha\beta}}{\rho_a\rho_b}+C_{ab}^{\alpha\beta}\right)\tilde{\nabla}_a^\beta W_{ab}$$
$$+\sum_{b=1}^N \frac{mb}{\rho_a\rho_b}(p_{wb}-p_{wa})\tilde{\nabla}_a^\alpha W_{ab} + g_a^\alpha \qquad (24)$$

Alternatively, if we keep the same form of the pore-water pressure approximation as derived in equation (22), and combine this approximation with equation (14) the following equation can be derived,

$$\ddot{u}_a^\alpha = \sum_{b=1}^N m_b\left(\frac{\sigma_a'^{\alpha\beta}}{\rho_a^2}+\frac{\sigma_b'^{\alpha\beta}}{\rho_b^2}+C_{ab}^{\alpha\beta}\right)\tilde{\nabla}_a^\beta W_{ab}$$
$$+\sum_{b=1}^N \frac{m_b}{\rho_a\rho_b}(p_{wb}-p_{wa})\tilde{\nabla}_a^\alpha W_{ab} + g^\alpha \qquad (25)$$

For homogenous ground density, equation (24) and (25) are identical. However, for non-homogenous ground density where there is a significant change in soil density between two adjacent soil layers, equation (24) could give more stable result. Therefore, equation (24) will be considered as the fundamental equations for further developments of SPH for saturated soil in problems such as: soil-water coupling, slope failure due to rainfall, liquefaction, etc. Finally, equations (24) can be resolved using the standard Leapfrog algorithm if the effective stress tensor is known. Thus, it is necessary to derive a constitutive relation for the effective stress tensor that is applicable in the SPH framework.

### IV. SOIL CONSTITUTIVE MODEL

A soil constitutive model describes behaviour of a soil via relationship between stress and strain. So far, a number of soil constitutive models have been developed to model different kind of soils such as: elastic, elasto-plastic, cam-clay, critical state soil models, etc., and these models have been successfully implemented into the FEM code. In term of SPH, any soil constitutive model can be also implemented into the SPH method using the similar framework in [7]. In this paper, for the sake of simplicity, soil has been assumed linearly elastic. The stress-strain relation of an elastic soil model can be easily derived using a generalized Hooke's law. Accordingly, the elastic strain rate tensor can be written as,

$$\dot{\varepsilon}^{\alpha\beta} = \frac{\dot{s}'^{\alpha\beta}}{2G}+\frac{1-2\upsilon}{3E}\dot{\sigma}'^{\gamma\gamma}\delta^{\alpha\beta} \qquad (26)$$

where $\dot{s}'^{\alpha\beta}$ is the deviatoric effective shear stress rate tensor; $G$ is the shear modulus, $E$ is Young's modulus, and $\upsilon$ is the Poisson's ratio. By solving equation (26) for $\dot{s}'^{\alpha\beta}$ and using the following relation,

$$\dot{\sigma}'^{\alpha\beta} = \dot{s}'^{\alpha\beta} + \tfrac{1}{3}\dot{\sigma}'^{\gamma\gamma}\delta^{\alpha\beta} \qquad (27)$$

The stress-strain relation for an elastic soil model at particle $a$ can be written as,

$$\dot{\sigma}_a'^{\alpha\beta} = 2G_a(\dot{\varepsilon}_a^{\alpha\beta}-\tfrac{1}{3}\dot{\varepsilon}_a^{\gamma\gamma}\delta^{\alpha\beta})+K_a\dot{\varepsilon}_a^{\gamma\gamma}\delta^{\alpha\beta} \qquad (28)$$

where $K_a$ is the elastic bulk modulus; and $\dot{\varepsilon}_a^{\alpha\beta}$ is the strain rate tensor at particle $a$ defined by,

$$\dot{\varepsilon}_a^{\alpha\beta} = \frac{1}{2}\left[\sum_{b=1}^{N}\frac{m_b}{\rho_b}(\dot{u}_b^{\alpha} - \dot{u}_a^{\alpha})\tilde{\nabla}_a^{\beta}W_{ab} + \sum_{b=1}^{N}\frac{m_b}{\rho_b}(\dot{u}_b^{\beta} - \dot{u}_a^{\beta})\tilde{\nabla}_a^{\alpha}W_{ab}\right] \quad (29)$$

In order to guarantee the independence of formulation from rigid body rotation, the Jaumann stress rate is adopted here in with the following constitutive equation as,

$$\dot{\sigma}'^{\alpha\beta}_a = 2G_a(\dot{\varepsilon}_a^{\alpha\beta} - \tfrac{1}{3}\dot{\varepsilon}_a^{\gamma\gamma}\delta_a^{\alpha\beta}) + K_a\dot{\varepsilon}_a^{\gamma\gamma}\delta_a^{\alpha\beta} \\ + \sigma'^{\alpha\gamma}_a\dot{\omega}_a^{\beta\gamma} + \sigma'^{\gamma\beta}_a\dot{\omega}_a^{\alpha\gamma} \quad (30)$$

where $\dot{\omega}_a^{\alpha\beta}$ are the spin rate tensors defined by,

$$\dot{\omega}_a^{\alpha\beta} = \frac{1}{2}\left[\sum_{b=1}^{N}\frac{m_b}{\rho_b}(\dot{u}_b^{\alpha} - \dot{u}_a^{\alpha})\tilde{\nabla}_a^{\beta}W_{ab} - \sum_{b=1}^{N}\frac{m_b}{\rho_b}(\dot{u}_b^{\beta} - \dot{u}_a^{\beta})\tilde{\nabla}_a^{\alpha}W_{ab}\right] \quad (31)$$

The above soil model requires three soil parameters, which are Young's modulus ($E$), Poisson's ratio ($\upsilon$), and soil density ($\rho$).

## V. INITIAL STRESS AND BOUNDARY CONDITIONS

### A. Initial Stress Condition

Many problems in geotechnical engineering require the specification of a set of initial stresses. These stresses, which are caused by gravity, represent the equilibrium state of the undisturbed soil body. Computations that employed inappropriate initial stresses will result in unrealistic predictions, which are very dangerous for geotechnical designs. Therefore, cares must be taken to correctly obtain the initial stress condition in advance of calculations.

Basically, there are two common methods that are often used to generate initial stress conditions in soil: $K_0$ method and gravity loading method. In the $K_0$ method, the vertical stress is calculated as a product between the unit weight of soil and its elevation, while the lateral stresses is a product between earth pressure coefficient $K_0$, which may be taken based on Jaky's formula (1-sin$\phi$), and the vertical stress. Although, this method is very simple it can only be used for horizontally layered geometries with a horizontal ground surface and horizontal underground water level. For soils with non-horizontal ground surface, the second gravity loading method is often employed where the initial stresses were created by applying soil self-weight in the fist calculation phase.

In SPH analysis, both methods described above can be employed, in an appropriate way, to obtain the initial stresses of a soil. However, additional cares must be taken in order to ensure that the initial stresses in soil represent an equilibrium state of a soil. In fact, as similar to other particle methods, very large velocity and stress fluctuations have been observed in SPH when suddenly applying load to a soil body. Such the fluctuations came from the fact that the SPH method was suffered from the zero-energy mode where both velocity and stress are interpolated at the same location [22]. In this paper, in order to damp-out such the fluctuations, a damping force has been added into the momentum equation during the stress loading phase in the following manner,

$$\ddot{u}_a^{\alpha} = \sum_{b=1}^{N} m_b\left(\frac{\sigma'^{\alpha\beta}_a + \sigma'^{\alpha\beta}_b}{\rho_a\rho_b} + C_{ab}^{\alpha\beta}\right)\tilde{\nabla}_a^{\beta}W_{ab} \\ + \sum_{b=1}^{N}\frac{m_b}{\rho_a\rho_b}(p_{wb} - p_{wa})\tilde{\nabla}_a^{\alpha}W_{ab} + D_a^{\alpha} + g_a^{\alpha} \quad (32)$$

where $D$ is the damping force per unit mass defined by,

$$D_a^{\alpha} = -c_d\dot{u}_a^{\alpha} \quad (33)$$

with $c_d$ is a damping coefficient.

The damping coefficient can be modeled by using Rayleigh damping and its alternatives. For the sake of simplicity, this paper employed the following damping coefficient [23],

$$c_d = \xi / dt \quad (34)$$

where $\xi$ is a non-dimensional damping coefficient. Our numerical tests showed that the effective damping coefficient should be chosen in range of $\xi$ = 0.001 - 0.005.

It must be kept in mind that the damping force employed in equation (32) should be only used for the purpose of obtaining the initial stresses distribution in a soil. On the other hand, when soil deformation analysis is started this force must be removed to avoid incorrect results caused by energy lost due to damping.

### B. Boundary Condition

There have been several methods developed to model solid boundary conditions in SPH such as: ghost particles to model the free-slip boundary conditions [24]; repulsive force boundary condition [23], which was the simplest free-slip boundary condition; no-slip condition for viscous fluid [25-26]; stress boundary condition [8]; etc. In this paper, most problems apply two types of boundary conditions: free-roller and full-fixity; the free-roller boundary condition is modelled using ghost particles [24], while the full-fixity one can only be modelled using the stress boundary method whereby virtual particles are used to model the solid boundary and an additional procedure assigns velocity and stress to these boundary particles [8].

## VI. NUMERICAL VALIDATIONS

In this section, validations of the proposed SPH formulation for saturated soil will be presented via some numerical tests. Role of the damping force and its effects are also discussed throughout.

## A. Horizontal ground surface: A Fully Submerged Soil Foundation Subjected to Gravity Loading

A plane-strain problem of saturated soil foundation submerged into water is considered in this section. The geometry and boundary conditions of this soil foundation are shown in Figure 2 where the boundary conditions have been restrained at the lateral boundaries and fixed in both directions at the bottom boundary. The material has been assumed isotropic linear elastic, with the following properties: $E = 15 \times 10^6$Pa, $\upsilon = 0.33$, $\gamma_{sat} = 20$kN/m$^3$, $\gamma_w = 9.81$kN/m$^3$. In SPH, the above soil foundation has been modelled using 3750 discrete particles with a smoothing length of 0.24m. Initially, all stress components of soil in the foundation were set to zero, the gravity loading, which includes self-weight loading and pore-water pressure loading, was then applied to the soil foundation, in a single increment, in order to obtain initial stresses distribution in soil. To avoid the stress fluctuation, which is caused by suddenly applying stress to the foundation, the damping force has been adopted with the damping coefficient taken to be $\xi = 0.002$. The accuracy of the proposed SPH formulation is evaluated by comparing the effective stresses at point A and point B to analytical solutions.

Figure 3 shows the contour plot of the total vertical stress distribution in the soil foundation. It can be seen that the conventional SPH formulation could not be applied to the current problem. Soil particles on the top foundation were expelled from the soil structure causing numerical instability problem. These soil particles behave exactly as explained in section III, i.e. when using the conventional SPH formulation, the extra pressure force has been introduced into the soil particles on the ground surface and pushing them upward. We have tried to resolve this problem by applying the same amount of the extra pressure force to soil particles on the ground surface to model the boundary condition between submerged soil and water, no improvements have been obtained. A more complex boundary condition may need to resolve this numerical instability problem which is very time consuming and requires a lot of effort. Contrarily, significant improvement has been obtained after adopting the new SPH formulation, which is straightforward. As for the accuracy of the proposed formulation, Figure 4 shows the comparisons of the effective vertical stresses at points A and B between SPH and theoretical solutions. It can be seen that the proposed SPH formulation with the kernel gradient correction agrees very well with the theoretical solutions. On the other hand, somewhat differences between the proposed SPH formulation without the kernel gradient correction and theoretical solutions have been observed. This result suggests that the kernel gradient correction could improve the accuracy of calculation in the current application. Furthermore, the matching result between the proposed formulation and theoretical solution also suggests that this formulation automatically satisfies the boundary condition between submerged soil and water, i.e. there is no need an effort to impose the hydrostatic water pressure due water reservoir to soil particles on the interface between submerged soil and water. This in turn could save much of the computational time which may need to search for particles on the interface and to assign pressure force to these particles.

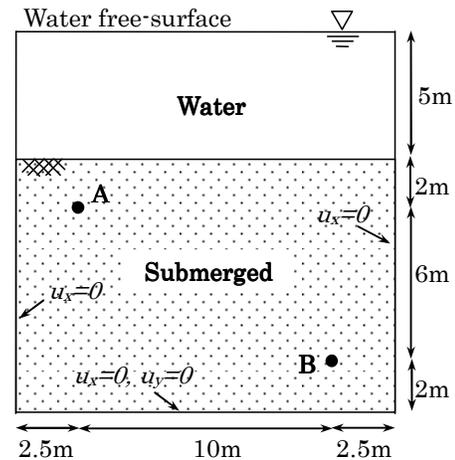

Figure 2. Geometry and boundary conditions of the soil foundation model.

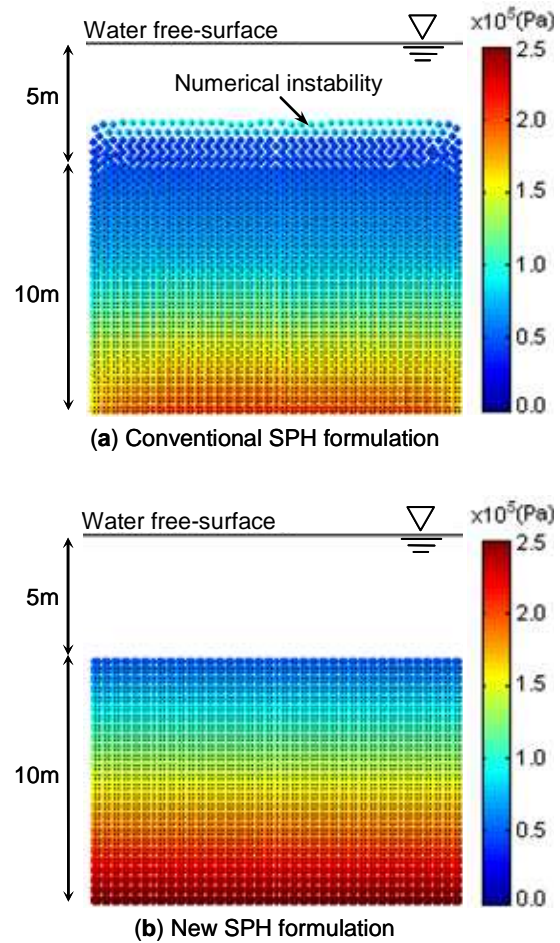

Figure 3. Initial stress distribution via the gravity loading procedure.

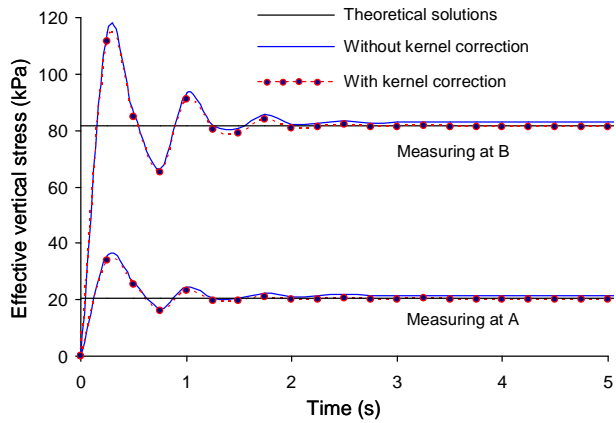

Figure 4. Development of the effective vertical stress at points A and B via the gravity loading procedure.

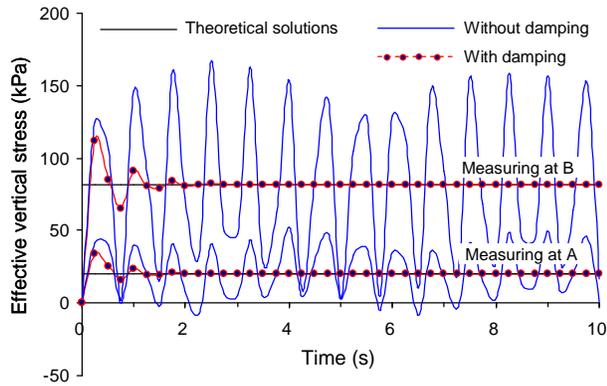

Figure 5. Effect of damping force on the development of the effective vertical stress at points A and B during the gravity loading procedure.

Regarding the role of the damping force in equation (32), Figure 5 shows the comparison of the results obtained by applying the damping force with those of without damping force. When the damping force was not employed, the stresses at points A and B were strongly fluctuated. This make us difficult to distribute the initial stress condition in soil foundation. On the other hand, by adopting the damping force with $\xi = 0.002$, the fluctuation was significantly reduced and competently removed after 2s. The magnitude of the stresses were not affected by the damping force and they were agree very well with the theoretical solutions. The above result suggests that the damping force could help to obtain the initial stress condition in soil.

### B. Non-horizontal ground surface: A Two-side Slope Embankment Subjected to Gravity Loading

Next, we will extend our test to a non-horizontal ground surface problem, which is commonly found in computational geomechanics. A two-side slope embankment geometry and boundary conditions considered herein are shown in Figure 6 with a stiffer slope on the left side. The material has been assumed isotropic linear elastic, with the following properties: $E$ = 15MPa, $\upsilon$ = 0.25, $\gamma_{sat}$ = 20kN/m$^3$, $\gamma_{unsat}$ = 18.6kN/m$^3$, $\gamma_w$ = 9.81kN/m$^3$. The current example is the typical problem which can not be modeled using the conventional SPH since both embankment foundations on the left and right sides are completely submerged into the water. Therefore, the proposed formulation with kernel gradient correction has been applied.

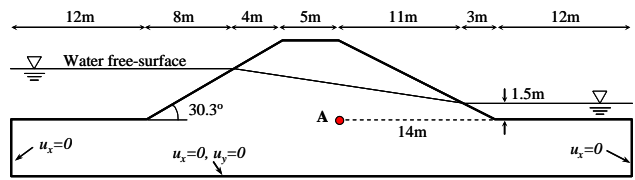

Figure 6. Two-side slope embankment model.

A total of 8454 particles have been used to represent the above embankment model with a smoothing length of 0.24m. Similar to the previous test, all stress components of the soil in the current embankment were initially set to zero. The gravity loading, which includes self-weight loading and pore-water pressure loading, was then applied to the embankment in a single increment and the damping force ($\xi$ = 0.002) was also adopted to remove the stress fluctuation. Results including contour plot of stress components and stress measure at point A are then validated with those obtained by the finite element method (FEM). The FEM code employed 15-noded triangular element and the updated Lagrangian formulation was also adopted in an attempt to capture large deformations of soil.

Figure 7 shows the comparison between SPH and FEM in term of the total vertical stress distribution in the embankment. The contour plot shows that result predicted by SPH agrees very well with FEM. Contrarily, soil particles on the both sides of the embankment foundation and slope toe are expelled from the top surface when adopting the conventional SPH formulation. As for the accuracy of the formulation in the current application, Figure 8 shows the comparison between SPH and FEM in term of the effective stresses measured at point **A**. Again, very good agreement was obtained. These results suggest that the proposed SPH formulation would be applied well to model the saturated soil.

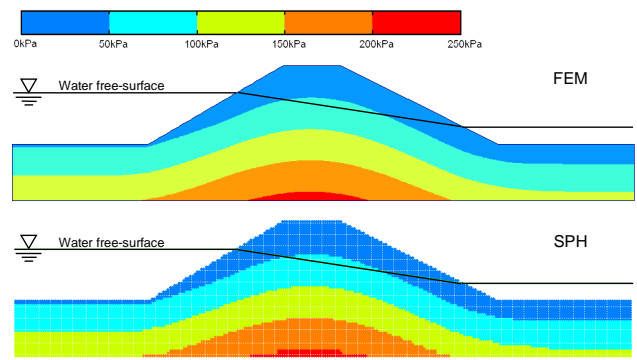

Figure 7. Comparsion of vertical stress ($\sigma_{yy}$) between SPH and FEM.

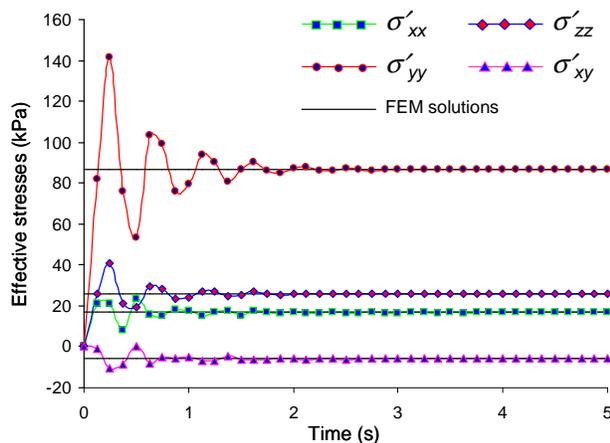

Figure 8. Comparsion of the effective stress components measured at point A between SPH and FEM.

## VII. Conclusion

A general SPH formulation with kernel gradient correction for dry and saturated soils has been proposed through out this paper. It is shown that the new formulation can easily remove the numerical instability caused by using the conventional SPH formulation when dealing with a fully submerged soil without additional efforts. The formulation is very robust and can be applied to a wide range of problems. Furthermore, the formulation automatically satisfied the boundary condition on the interface between submerged soil and water, thereby significantly saving the computational time. For the purpose of generalizing SPH to model large deformation and post-failure of geomaterials, the proposed formulation significantly contribute to this progress.

## Acknowledgement

This research is supported by the Japan Society for the Promotion of Science (JSPS) through the JSPS-Postdoctoral Fellowship for Foreign Researchers (P-10072) and the Research Grants for JSPS Postdoctoral Fellowships.